\documentclass[preprint,showpacs,preprintnumbers,amsmath,amssymb]{revtex4}

\usepackage{graphicx}% Include figure files
\usepackage{dcolumn}% Align table columns on decimal point
\usepackage{bm}% bold math
\usepackage{color}%' 'Æ'Å'¯'·
\begin{document}

\title{Critical Velocity in a Bose Gas in a Moving Optical Lattice at Finite Temperatures}

\author{Emiko Arahata}
\author{Tetsuro Nikuni}%
\affiliation{%
Department physics, Faculty of science, Tokyo University of Science, \\
1-3 Kagurazaka, Shinjuku-ku, Tokyo 162-8601, Japan}%
\date{\today}
\begin{abstract}We study the critical velocity of a Bose-condensed gas in a moving one-dimensional (1D) optical lattice potential at finite temperatures. 
Solving the Gross-Pitaeavskii equation and the Bogoliubov equations, 
within the Popov approximation, we calculate the Bogoliubov excitations with varying lattice velocity.
From the condition of the negative excitation energy, we determine the critical velocity as a function of the lattice depth and the temperature. We find that the critical velocity decreases rapidly with increasing the temperature; this result is consistent with the experimental observations. Moreover, the critical velocity shows a rapid decrease with increasing lattice depth. This tendency is much more significant than in the previous works ignoring the effect of thermal excitations in the radial direction. 
\end{abstract}
\maketitle
\section{Introduction}
The occurrence of energetic and dynamical instabilities in
a Bose-Einstein condensate (BEC) moving in a periodic
(optical lattice) potential is an interesting problem from the conceptual
viewpoint, since it involves basic properties of superfluidity.
Experimentally, energetic instabilities of BECs moving
through optical lattices have been observed both in the
weak lattice and tight-binding regimes \cite{E_Flo_Instability,E_T_Flo_density_dispative,E_NIST_Dynamical_Instability}.
According to Landau's argument \cite{B_Pethich_Smith}, 
the energetic instability of the superfluid state is attributed to the appearance of negative excitation energy (which is known as Landau instability).  Once a superfluid velocity is beyond its unique critical velocity, the current suffers friction leading to the
decay of superfluidity. 
We note that the actual occurrence of energetic instability requires some dissipative mechanism, which leads the system to a lower energy state. Some theoretical studies 
showed that the thermal component plays a crucial role, receiving the energy emitted by the condensate during the breakdown process \cite{Konabe,iigaya}. Experimentally, energetic instability is only observed at finite temperatures, consistent with the theory of Refs \cite{Konabe,iigaya}. Thus, in order to provide a quantitative account for the critical velocity relevant to the experimental data, it is important to consider thermal component at finite temperatures.
However, most theoretical studies on the Landau instability of a Bose gas in an optical lattice used the zero-temperature GP equation \cite{T_Smerzi_sound_BEC,T_Menotti_NJP5}. Moreover, they concentrated on the first Bloch band using the Bose-Hubbard model, and have ignored the effect of radial excitations. It will turn out that one should include thermal excitation in the radial direction to obtain quantitative results. \par
In this paper, we study the critical velocity of current carrying condensate in a 1D optical lattice at finite temperatures. For this purpose, we calculate the Bogoliubov excitation energy in the framework of Hatrree-Fock-Bogoliubov-popov (HFB-Popov) approximation in a periodic potential, with explicitly including the effect of the radial excitations.
In Sec.~II, we derive a quasi-1D model of the Gross-Pitaeavskii equation and the Bogoliubov equations for the Bose gas moving in a 1D optical lattice. Using the HFB-Popov approximation, we solve these equations to calculate the Bogoliubov excitation spectrum. 
In Sec.~III, we calculate the critical velocity of current carrying condensate in a optical lattice, which will be determined from the condition of the negative excitation energy. We obtain the critical velocity as a function of the lattice depth with a fixed temperature, and compare it with the experimental data \cite{E_Flo_Instability}. 
The magnitude of the critical velocity is found to be in reasonable agreement of the experimental data \cite{E_Flo_Instability}. We will also show that the critical velocity decreases with increasing lattice depth, which is consistent with the experimental result \cite{E_Flo_Instability}. We also calculate the temperature dependence of the critical velocity with a fixed lattice depth. The critical velocity will be found to drop very rapidly with increasing temperature.
\section{Quasi 1D modeling of a current carrying condensate}
We consider a Bose condensed gas in a combined potential of highly elongated harmonic trap and 1D optical lattice.
Our system is described by the following Hamiltonian :
\begin{eqnarray}
\hat{H}=\int d\textbf{r}\Bigg\{\hat{\psi}^\dagger (\textbf{r})\left[-\frac{\hbar^2}{2m}\nabla ^2+V_{\rm{ext}}(\textbf{r})\right]\hat{\psi }(\textbf{r})\nonumber \\ 
+\frac{g}{2}\hat{\psi}^\dagger(\textbf{r})\hat{\psi}^\dagger (\textbf{r})\hat{\psi }(\textbf{r})\hat{\psi }(\textbf{r})\Bigg\},\label{H}
\end{eqnarray}
where $g=\frac{4\pi\hbar^2a}{m}$ is the coupling constant determined by the $s$-wave scattering length $a$. 
The external potential $V_{\rm{ext}}$ is given by
$V_{\rm{ext}}(\textbf{r})=V_{\rm{trap}}(\textbf{r})+V_{\rm{op}}(z)$.
More explicitly, we consider a cylindrical condensate, which is radially confined by the harmonic potential
$V_{\rm{trap}}(\textbf{r})=\frac{m}{2}\left[\omega _\bot ^2(x^2+y^2)\right]$, and is subject to the periodic potential
$V_{\rm{op}}(z)=sE_{\rm{R}}\cos^2(\frac{G}{2}z) $.
 Here $s$ is a dimensionless parameter describing the intensity of the laser beam creating the 1D lattice in units of the recoil energy $E_{\rm{R}}\equiv\frac{\hbar^2k^2}{2m}$, where $\frac{G}{2}=\frac{2\pi}{\lambda}$ is fixed by the wavelength $\lambda$ of the laser beam. 
In this paper, we neglect the harmonic trap potential along the $z$-direction. we consider a highly anisotropic cigar-shaped harmonic trap potential. In order to take into account this quasi-1D situation, we expand the field operator
in terms of the radial wave function 
\begin{eqnarray}\hat{\psi }(\textbf{r})=\sum _{\alpha }\hat{\psi }_{\alpha}(z)\phi _\alpha (x,y),\label{xyz}\end{eqnarray} 
where $\phi_\alpha(x,y) $ is the eigenfunction of the radial part of the single-particle Hamiltonian \cite{cd06,cd106}. 
\begin{eqnarray} 
&&\left[-\frac{\hbar^2}{2m}\nabla _{\bot}^2+\frac{m}{2}\omega _\bot ^2(x^2+y^2)\right]\psi _\alpha (x,y)=\epsilon _\alpha\psi _\alpha (x,y) ,\label{xy}
\end{eqnarray} 
which satisfy the orthonormality condition $ \int dxdy \psi ^\ast _\alpha (x,y)\psi _\beta  (x,y)=\delta _{\alpha \beta}$.
Here $\alpha=(n_x,n_y)$ is the index of the single-particle state with the eigenvalue $\epsilon_{(n_x,n_y)}=\hbar\omega_\bot(n_x+n_y+1)$, where $\omega_\bot$ is the trap frequency is the radial direction. Inserting Eq.~(\ref{xyz}) into Eq.~(\ref{H}) and using Eq.~(\ref{xy}), we obtain 
\begin{eqnarray}
\hat{H}=\sum _\alpha \int dz \hat{\psi }_{\alpha}(z)\left[-\frac{\hbar^2}{2m}\frac{\partial ^2}{\partial z^2}+V_{op}(z)+\epsilon _\alpha \right]\hat{\psi }_{\alpha}(z)\nonumber\\
+\sum _{\alpha  \alpha ^\prime \beta \beta^\prime}\frac{g_{\alpha  \alpha ^\prime \beta \beta^\prime}}{2}\int dz \hat{\psi }_{\alpha}^\dagger \hat{\psi }_{\beta}^\dagger \hat{\psi }_{\beta ^\prime}\hat{\psi }_{\alpha ^\prime}. \label{H_k}
\end{eqnarray}
The renormalized coupling constant is defined by
$
g_{\alpha  \alpha ^\prime \beta \beta^\prime}\equiv g\int dxdy \phi _{\alpha}^\ast \phi _{\beta }^\ast \phi _{\beta ^\prime}\phi _{\alpha ^\prime}
$. 
Following the procedure described in Ref.~\cite{T_HFB-popov_Griffin}, we separate out the condensate wavefunction from the field operator as 
$
\hat{\psi}_\alpha=\langle\hat{\psi}_\alpha\rangle+\tilde{\psi}_\alpha \equiv \Phi_\alpha+\tilde{\psi}_\alpha 
$,
where $ \Phi_\alpha=\langle\hat{\psi}_\alpha\rangle$ is the condensate wavefunction and $\tilde{\psi}_\alpha$ is the noncondensate field operator.
The Popov approximation neglects the anomalous correlation  $\left\langle \tilde{\psi}\tilde{\psi}  \right\rangle$ \cite{T_HFB-popov_Griffin}. 
Within the HFB-Popov approximation, we obtain the generalized Gross-Pitaevskii (GP) equation \cite{cd06,cd106},
\begin{eqnarray}
&&\mu \Phi _\alpha=\left[-\frac{\hbar^2}{2m}\frac{\partial ^2}{\partial z^2}+V_{\rm{op}}+\epsilon _\alpha \right]\Phi _\alpha+\sum _{\alpha ^\prime \beta \beta^\prime}g_{\alpha  \alpha ^\prime \beta \beta^\prime}\left(\Phi^\ast _{\beta}\Phi _{\beta^\prime}\Phi _{\alpha^\prime}+2\Phi _{\beta^\prime} \left\langle\tilde{\psi}_{\beta}^\dag \tilde{\psi}_{\alpha ^\prime}\right\rangle  \right).
\label{GP}
\end{eqnarray}
Solving the GP equation (\ref{GP}) at $T=0$ (setting $\left\langle \tilde{\psi}^\dagger\tilde{\psi}  \right\rangle=0$ ) using the trap frequencies relevant to the experiment \cite{E_Flo_Instability} (see the paragraph below Eq.~(\ref{H_k}) ), we find that \textcolor{blue}{$\mu\ll \hbar\omega_\bot$ and} $|\Phi _\alpha|^2/|\Phi _0|^2\ll10^{-6}$ ($\alpha\neq0$), where we have denoted
the lowest radial state as $\alpha=0=(0,0)$.  Thus the
contribution from higher radial modes to the condensate
wave function is negligible small.  For this reason, we will
henceforth approximate $\Phi_\alpha\approx\Phi\delta_{\alpha,0}$. 
Taking the usual Bogoliubov transformations for the noncondensate,
$
\tilde{\psi}_\alpha(z,t)=\sum_{jk} \left[u_{jk\alpha}\hat{\alpha}_{jk} -v_{jk\alpha}^*\hat{\alpha}^\dag_{jk}   \right]$,
 we obtain the coupled Bogoliubov equations,
 \begin{eqnarray}
 &\hat{L}_\alpha u_{j \alpha}+\sum_{\alpha^\prime} \Biggl[ \left( 2g^\alpha_{\alpha^\prime} n_0 +g_{\alpha\alpha^\prime \beta \beta^\prime}\tilde{n}_{\beta\beta^\prime} \right) u_{j \alpha^\prime}-g^\alpha_{\alpha^\prime} n_0 v_{j \alpha^\prime} \Biggr]
=E_j u_{j \alpha},\nonumber \\
& \hat{L}_\alpha v_{j \alpha}+\sum_{\alpha^\prime} \Biggl[ \left( 2g^\alpha_{\alpha^\prime} n_0 +g_{\alpha\alpha^\prime \beta \beta^\prime}\tilde{n}_{\beta\beta^\prime}\right) v_{j \alpha^\prime}-g^\alpha_{\alpha^\prime} n_0 u_{j \alpha^\prime} \Biggr]
=-E_j u_{j \alpha},\label{Bog}
 \end{eqnarray}
 where we have introduce the operator 
 $\hat{L}_\alpha\equiv-\frac{\hbar^2}{2m}\frac{\partial^2 }{\partial z^2}+V_{\rm{op}}(z)+\epsilon _\alpha-\mu
$.
 We have also introduced the simplified notations 
 $n_0(z)=|\Phi(z)|^2,
   g^\alpha_{\alpha^\prime}=g_{\alpha\alpha^\prime00},
 \tilde{n}_{\beta\beta^\prime}=\left\langle\tilde{\psi}^\dag_\beta\tilde{\psi}_{\beta^\prime}  \right\rangle $. As noted above, we only include the  lowest mode ($\alpha=0$) in the condensation wavefunction $\Phi$.  
 Sums over the repeated indices $\beta, \beta^\prime$ are implied in Eq.~(\ref{Bog}).
 These equations define the quasi-particle excitation energies $E_j$ and the quasi-particle amplitudes $u_{j\alpha}$ and $v_{j\alpha}$. 
 Using the solutions of Eq.~(\ref{Bog}), one can obtain the noncondensate density from 
$\tilde{n}=\sum_{\alpha}\tilde{n}_{\alpha \alpha}$,
where 
$
\tilde{n}_{\alpha \beta}=\sum_{j}\biggl[\left(u_{j\alpha}u_{j\beta}+v_{j\alpha}v_{j\beta}\right)N(E_j)+v_{j\alpha}v_{j\beta}\biggr]$
with $N(E_j)=1/\left[ \exp(\beta E_j )-1\right]$.
\par
We expand the condensate wavefunction in terms of the reciprocal lattice vector  $G$, $\Phi(z)=\sum_n e^{i(q+nG)z}C_n$. Here we consider the situation where the condensate is moving with the velocity $\hbar q$.
We also expand the Bogoliubov quasiparticle amplitudes $u_{k\alpha}$ and $v_{k\alpha}$ in terms of the reciprocal lattice vector,
$
u_{k\alpha}=\sum_n u_{kn}^\alpha e^{i(nG+k)z},
v_{k\alpha}=\sum_n v_{kn}^\alpha e^{i(nG+k)z},
$
where $k$ is the Bloch wavevector (or quasimomentum) of
excitations. Then, GP equation and Bogoliubov equations become 
\begin{eqnarray}
&& \Biggl[\left(\frac{\hbar^2}{2m}\left(q+nG\right)^2+\epsilon_0-\mu+\frac{sE_{\rm{R}}}{2}\right)C_n
+\frac{sE_{\rm{R}}}{4}\left(C_{n+1}+C_{n-1}\right)
+g\sum_{mm^\prime}C^\ast_{m^\prime}C_{m}C_{n+m^\prime-m}
\nonumber\\ 
&&+2\sum_{\alpha^\prime \beta}g_{\alpha^\prime \beta}\sum_{jk ll^\prime}\Big\{\left(u_{jkl^\prime}^{\alpha^\prime}u_{jkl}^{\ast\beta}C_{n+l-l^\prime}
+v_{jkl^\prime}^{\ast\alpha^\prime} v_{jkl}^\beta C_{n+l^\prime-l}  \right)N^0_{jk}
+v_{jkl^\prime}^{\ast\alpha^\prime} v_{jkl}^\beta C_{n+l^\prime-l} \Big\}
\Biggr],\label{GPk}
\end{eqnarray}
\begin{eqnarray}
&&\Bigg\{\frac{\hbar^2}{2m}\left(q\pm k+nG\right)^2+\frac{sE_{\rm{R}}}{2}-\mu+\epsilon_\alpha\Bigg\}
\left ( 
\begin{array}{c}
u_{jkn}^\alpha\\
v_{jkn}^\alpha
\end{array}
\right)
+\frac{sE_{\rm{R}}}{4}\Bigg\{
\left ( 
\begin{array}{c}
u_{jkn+1}^\alpha\\
v_{jkn+1}^\alpha
\end{array}
\right)
+\left ( 
\begin{array}{c}
u_{jkn-1}^\alpha\\
v_{jkn-1}^\alpha
\end{array}
\right)\Bigg\}\nonumber\\
&&
+\sum_{\alpha^\prime}\Biggl[2g_{\alpha\alpha^\prime}\sum_{mm^\prime}C^\ast_{m^\prime}C_{m}
\left ( 
\begin{array}{c}
u_{jkn+m^\prime-m}^{\alpha^\prime}\\
v_{jkn+m^\prime-m}^{\alpha^\prime}
\end{array}
\right)\nonumber\\
&&
+\sum_{\beta\beta^\prime }2g_{\beta\beta^\prime}\sum_{jk ll^\prime}\Bigg\{u_{jkl}^{\beta}u_{jkl^\prime}^{\ast\beta^\prime}N^0_{jk}\left ( 
\begin{array}{c}
u_{jkn+l^\prime-l}^{\alpha^\prime}\\
v_{jkn+l^\prime-l}^{\alpha^\prime}
\end{array}
\right)
+v_{jkl}^{\ast\beta} v_{jkl^\prime}^{\beta^\prime}\left(N^0_{jk}+1\right) \left ( 
\begin{array}{c}
u_{jkn+l-l^\prime}^{\alpha^\prime}\\
v_{jkn+l-l^\prime}^{\alpha^\prime}
\end{array}
\right)\Bigg\}
\Biggr]\nonumber\\
&&-\sum_{\beta}g_{\alpha\beta}\sum_{mm^\prime}
\left ( 
\begin{array}{cc}
0 & C_{m^\prime}C_{m}\\
C^\ast_{m^\prime}C^\ast_{m} & 0
\end{array}
\right)
\left ( 
\begin{array}{c}
u_{jk n+m^\prime+m}^{\beta}\\
v_{jk n-m^\prime-m}^{\beta}
\end{array}
\right)=\left ( 
\begin{array}{cc}
E_j & 0\\
0 & -E_j
\end{array}
\right)
\left ( 
\begin{array}{c}
u_{jkn}^\alpha\\
v_{jkn}^\alpha
\end{array}
\right).
\label{Bogk}
\end{eqnarray}
The dispersion relation for the condensate moving through the optical lattice is obtained by solving the Bogoliubov equations in Eq.~(\ref{Bogk}). Solving the coupled equations (\ref{GPk}) and (\ref{Bogk}), we self-consistently determine the excitations spectrum $E_j$ and the condensate fraction at finite temperatures.
Our calculation procedure closely follows that summarized in Refs.~\cite{cd06,cd106}.
Throughout this paper we use the following parameters of the experiment of Ref.~\cite{E_Flo_Instability}: $m(^{87}$Rb)=1.44$\times 10^{-25}$ kg, $\omega_z/2 \pi= 9.0$ Hz, $\omega_{\bot}/2\pi= 92$ Hz, scattering length $a=5.82$ nm and the wavelength of the optical lattice $\lambda$=795 nm. We fixed the number of atoms per lattice site as $\int_{-d/2}^{d/2} n dz=200$. \par
\begin{figure}[htbp]
\centerline{\includegraphics[height=2.0in]{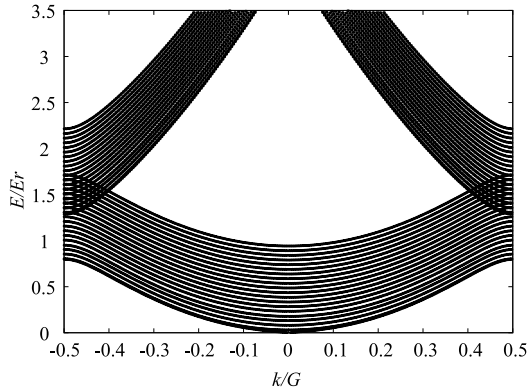}}%zu
 \caption{Excitation spectrum of a cylindrical condensate at rest in
the optical lattice (quasimomentum of the condensate $q$=0) as a
function of the quasimomentum $k$ of the excitations, for lattice depth
$s=1$. The first two Bloch bands are shown and for each band
we plot the first 20 radial branches. The lowest branch in each band
corresponds to axial excitations with no radial nodes.}
  \label{fig:Ene20}
\end{figure}

 
In Fig.~\ref{fig:Ene20} we show an excitation spectrum for $s=1$, $T=20$nK ($\simeq 0.14T_{\rm{c}}$) and $q=0$. The $q=0$ case corresponds to the condensate at rest. We only show the first two Bloch bands ($j$ =1 and 2). For each band, we plot the first 20 radial branches $\alpha =0,1,2, . . . ,19$. This spectrum is consistent with Modugno's results for $T=0$ \cite{T_radial_instability_Modugo}.

\begin{figure}[htbp]
\centerline{\includegraphics[height=2.0in]{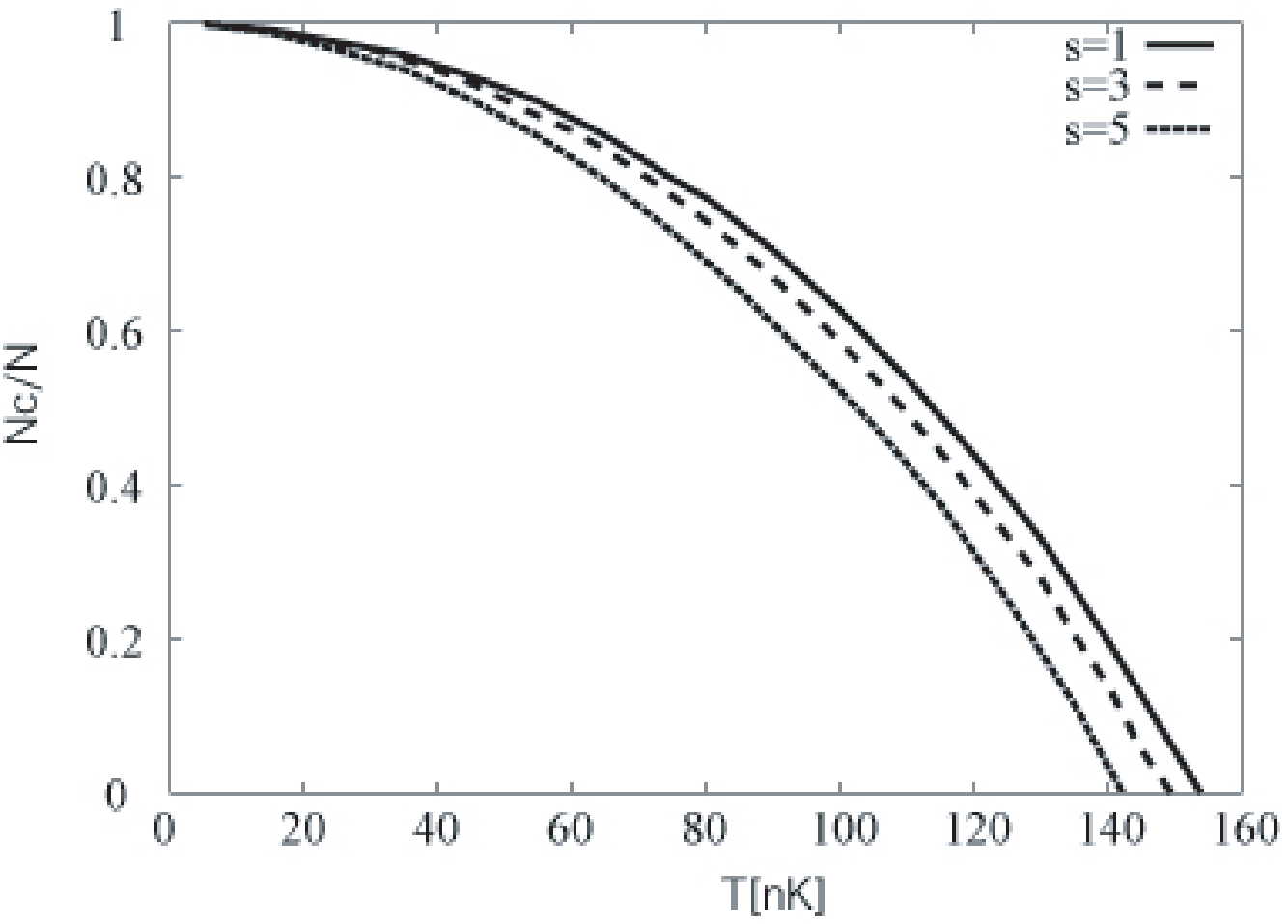}}%zu
 \caption{The condensate fraction as a function of temperature for different values of the lattice depth $s= 1, 3, 5$ }
  \label{fig:Nc0}
\end{figure}

In Fig.~\ref{fig:Nc0}, we plot the condensate fraction $N_{\rm{c}}/$$N$ as a function of the temperature for various values of the lattice depth $s$. 
In the actual calculation, we must introduce cutoff in the summation over the radial modes($\alpha$) and the reciprocal lattice vector($n$). We chose sufficiently large cutoff so that the results are cutoff-independent. It is crucial here that the cutoff is strongly temperature-dependent, because the thermal occupation of elementary excitations strongly depends on temperature. \par
As shown in Fig.~\ref{fig:Nc0}, the transition temperature deceases with increasing lattice depth.
In this paper, we focus on this weak lattice region $s<5$. In a stronger lattice $s>10$, one can use the tight binding approximation \cite{T_Blochband_Kramer_EPJD}. In contrast, a full HFB-Popov formalism presented in this paper must be used for a shallow lattice potential, say, $s<5$. In this paper, we only present the numerical results for $s<5$, but we expect the results to smoothly continue for $s>5$.
\section{The critical velocity of current carrying condensate}
\begin{figure}[htbp]
\centerline{\includegraphics[height=2.0in]{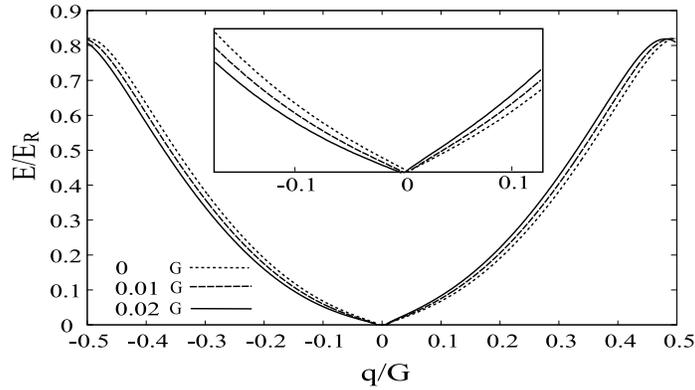}}%zu
 \caption{The excitation spectrum of the current carrying condensate with a fixed lattice depth $s$=1 and temperature $T=20$nK($\simeq 0.14T_{\rm{c}}$), for different values of the condensate momentum $q$ ($q/G$=0, 0.01, 0.02). We only plot the lowest radial branch (corresponding to axial excitations with no radial nodes).}
  \label{fig:Ene_q}
\end{figure}
\begin{figure}[htbp]
\begin{minipage}{16pc}
\includegraphics[width=16pc]{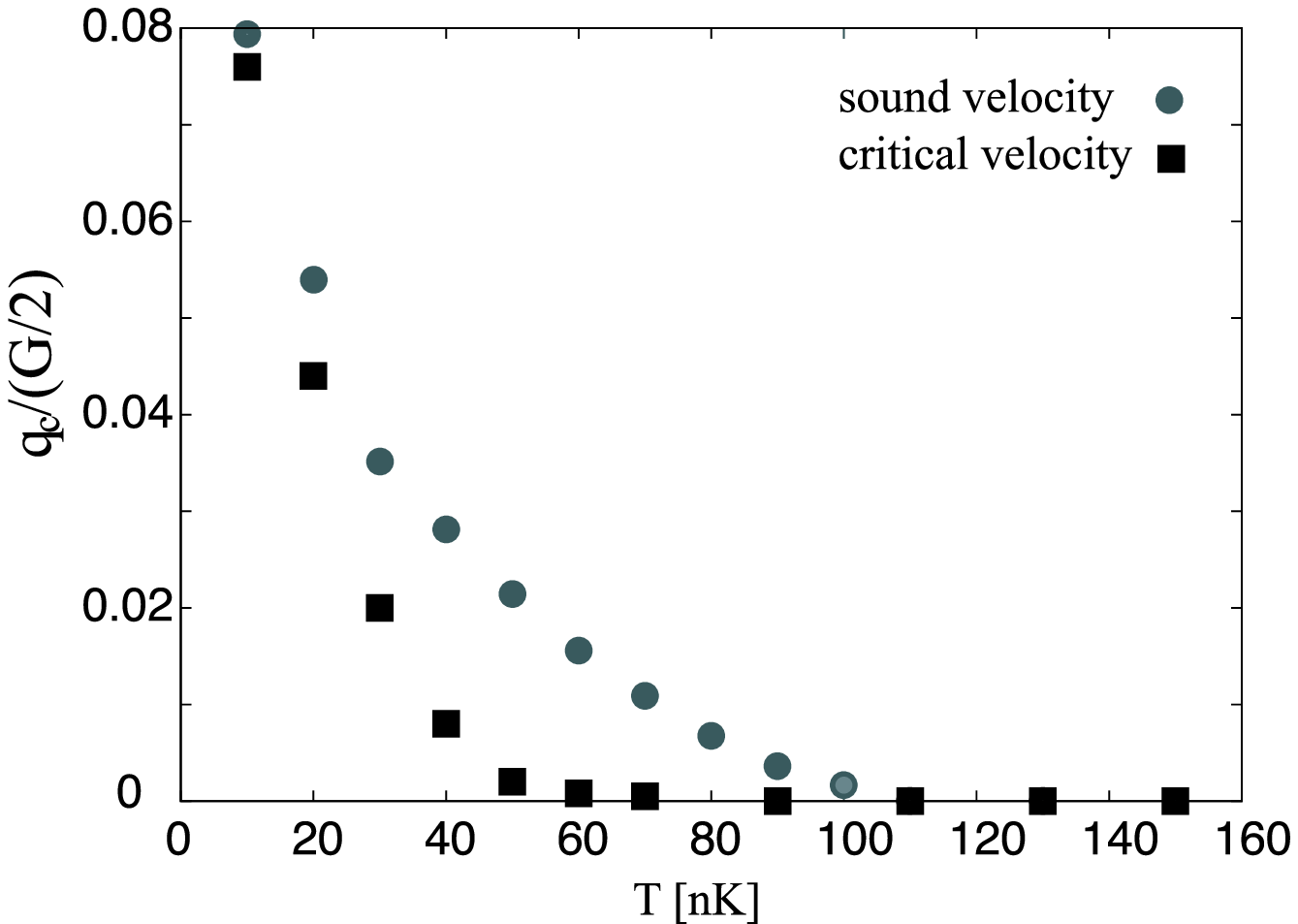}
\caption{\label{fig:CSvs1}Temperature dependence of Bogoliubov sound velocity ($q=0$) and the critical velocity with a fixed lattice depth $s$=1.}
\end{minipage}\hspace{2pc}%
\begin{minipage}{16pc}
\includegraphics[width=16pc]{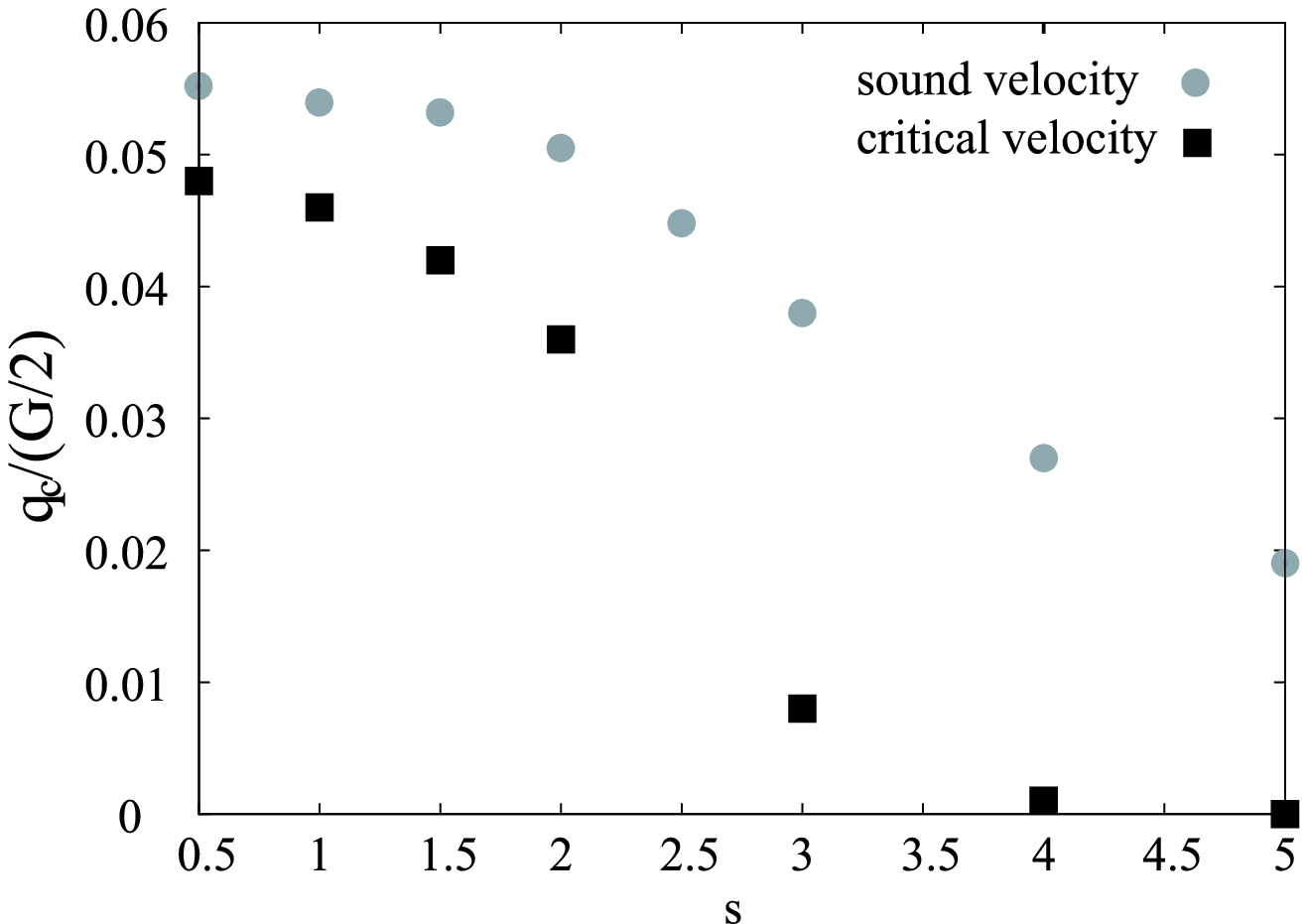} % 14pc
\caption{\label{fig:CSvT20}Dependence of Bogoliubov sound velocity and the critical velocity on the lattice depth $s$ with a fixed temperature $T=20$nK.}
\end{minipage} 
\end{figure}

We now consider the excitation spectra of the lowest phononic branch ($\alpha=0$) for $s=1$ at $T$=20nK($\simeq 0.14T_{\rm{c}}$) for different values of the condensate velocity (Fig.~\ref{fig:Ene_q}). By increasing $q$, the  dispersion law becomes asymmetric; the slope of the phononic branch decreases for the mode propagating opposite to the condensate current ($k<0$). By further increasing $q$, the slope of the spectrum at $k\rightarrow-0$ becomes zero at a certain critical value $q_{\rm{c}}$. Beyond $q_{\rm{c}}$, low energy excitations become negative. Therefore, superfluidity breaks down at $q_{\rm{c}}$. For comparison, we also calculate the Bogoliubov sound velocity $c$ for $q=0$ for different values of the temperatures $T$.
We obtained sound velocity $c$ from the slope of the Bogoliubov excitation spectrum at $k\rightarrow-0$.
In Fig.~\ref{fig:CSvs1}, we plot the temperature dependence of the critical velocity $q_{\rm{c}}$ and the sound velocity $c$ with fixing the lattice depth as $s=1$.
 It is clear that critical velocity is always lower than the sound velocity. We also find that critical velocity decreases rapidly with increasing temperature. The reduction of the critical velocity is more significant than that of the sound velocity. The reduction of both these velocities are due to the reduction of the condensate fraction with increasing temperature, which is larger for the condensate with a finite $q$ than the condensate at rest. This tendency is much more significant than in the previous works ignoring the effect of thermal excitations in the radial direction.  
\par We now compare our result with the experiment of Ref.~\cite{E_Flo_Instability}.
In the experiment with $35\%$ thermal fraction, a strong reduction of the number of atoms was observed even with an extremely low velocity, which indicates that critical velocity at finite $T$ is much lower than the sound velocity. 
Ref.~\cite{E_Flo_Instability} attributed this low critical velocity to the harmonic confinement in the $z$ direction. In fact, Ref.~\cite{E_Cv_crossover} pointed out that the lowering of both the critical velocity and sound velocity is caused by inhomogeneous density profile of the  harmonically trapped BEC. Another possibility of the observed extremely low critical velocity is a sudden onset of dissipation in the low density tail of the condensate cloud \cite{E_Cv_crossover,danshita}. On the other hand, from our calculation, the temperature corresponding to $35\%$ thermal fraction is estimated as $T\simeq 70.5$nk. As shown in Fig.~\ref{fig:CSvs1} at this temperature, the critical velocity has a strong temperature dependence, and the critical velocity at $70$nK is much lower than the sound velocity, i.e. $\hbar q_c/c(T=0) \ll 1$. Therefore, even ignoring the trap confinement in the $z$ direction, we find that the critical velocity becomes extremely low at the temperature relevant to the experiment \cite{E_Flo_Instability}. In any event, in order to make quantitative comparison between our results and experiments, it will be necessary to remove the effect of inhomogeneous density of the harmonically trapped BEC (for example, using the technique used in the experiment of Ref.~\cite{E_Cv_crossover} ). We note that it is crucial to include the thermal excitations in the radial direction in order to obtain the strong reduction of the critical velocity. 
\par In Fig.~\ref{fig:CSvT20}, we plot the Bogoliubov sound velocity $c$ and the critical velocity $q_{\rm{c}}$ as a function of lattice depth $s$ with fixing the temperature as $T=20$nK. We find that the critical velocity rapidly decreases with increasing lattice depth $s$. Fig.~\ref{fig:CSvT20} also shows that the critical velocity is always lower than the sound velocity $c$. 
The reductions of both the critical and sound velocities are due to the increase of the effective mass $m^\ast$ with increasing lattice depth. As shown in Refs.~\cite{T_m_PRA67,T_Blochband_Kramer_EPJD,T_Menotti_NJP5,Konabe}, $m^\ast$ depends on the condensate  momentum, and is larger for finite $q$. This gives rise to the significant change in the critical velocity.

\section{Conclusion}
In this paper, we studied the temperature dependence of the critical velocity of current carrying condensate in a 1D optical lattice, with explicitly including the effect of the radial excitations. In the trap geometry we considered, the condensate wavefunction can be treated only with the lowest radial mode, but the radial excitations for thermal cloud cannot be neglected. \par 
Within the Popov approximation, we calculated the temperature dependence of the Bogoliubov excitations with varying lattice velocity.
From the condition of the negative excitation energy, we determined the critical velocity as a function of the lattice depth and the temperature. 
For comparison, we also calculated the sound velocity of condensate at rest in a 1D optical lattice potentials. As shown in Fig.~\ref{fig:CSvs1}, the critical velocity rapidly decrease with increasing temperature, much faster than the sound velocity. The difference between the sound velocity and critical velocity was observed in the experiment of Ref.~\cite{E_Flo_Instability}.
Although the direct comparison of our results with the experimental data is difficult because of inhomogeneous density profile in the $z$ direction, the critical velocity is shown to be strongly reduced at the temperature relevant to the experiment.
\section*{ACKNOWLEDGMENTS}
We thank S.konabe and I. Danshita for valuable comments.
This research was support by Academic Frontier Project (2005) of MEXT.
%\bibliographystyle{apsrev}
%\bibliography{BEC.bib}

\end{document}